# Development of Wearable Systems for Ubiquitous Healthcare Service Provisioning

Ogunduyile, O.O[a*], Olugbara O.O[b], Lall M[c]

[a,c]Department of Computer Science, Tshwane University of Technology, Pretoria, South Africa
[b]Department of Information Technology, Durban University of Technology, Durban South Africa

**Abstract**

This paper reports on the development of a wearable system using wireless biomedical sensors for ubiquitous healthcare service provisioning. The prototype system is developed to address current healthcare challenges such as increasing cost of services, inability to access diverse services, low quality services and increasing population of elderly as experienced globally. The biomedical sensors proactively collect physiological data of remote patients to recommend diagnostic services. The prototype system is designed to monitor oxygen saturation level ($SpO_2$), Heart Rate (HR), activity and location of the elderly. Physiological data collected are uploaded to a Health Server (HS) via GPRS/Internet for analysis.

## 1. Introduction

Recent advancements in telemedicine technologies have led to the development of cost-effective wearable systems, which when fully implemented can be used in monitoring a person's physiological data ubiquitously. These systems provide a unique opportunity to address existing healthcare challenges of low quality services, inability to access various services, increasing costs of services and increasing population of elderly as experienced globally [1, 2]. A wearable system is important in ubiquitous healthcare and is characterised by the deployment of biomedical sensors around human body to proactively collect physiological data and transmit them wirelessly to a base node for processing [3]. There are existing studies on wearable systems for ubiquitous healthcare monitoring [4-10]. In particular, a Ubiquitous Healthcare System (UHS) was developed, which consists of vital signs devices and environment sensor devices to acquire context information to monitor and manage health status of patients anytime anywhere [4]. A wearable health system using wireless biomedical sensors for patient monitoring is presented [5]. The first level of the system consists of biomedical sensors, second level provides personal server and third level provides healthcare servers and related services. A new grid enabled framework, called gFrame is proposed for uubiquitous healthcare service provisioning [1]. Their work integrated e-healthcare technologies with Mobile Dynamic Virtual Communities (MDVC) into the healthgrid for cost-effective, quality and ubiquitous healthcare service provisioning. In spite of existing studies on wearable systems for ubiquitous healthcare, efficacious provisioning of healthcare services continues to remain a challenge in a wearable system ubiquitous health environment.

---

* Corresponding author.
*E-mail address*: gbengaroberts@yahoo.com

This study reports on life prototype implementation of a Wearable Ubiquitous Healthcare System (WUHS), which is designed for monitoring physiological data of the elderly. This is carried out using integrated wireless biomedical sensors (hereafter referred to as biomedical sensors) such as accelerometer and pulse oximeter ($SpO_2$). The accelerometer is used to monitor a user's body movement to determine his/her activity such as resting, walking, running and falling. The $SpO_2$ is used to measure blood-oxygen saturation levels ($SpO_2$) and Heart Rate (HR) [6]. The biomedical sensors are implemented on arduino fio platforms to collect physiological data and transmit them wirelessly to Intelligent Base Node (IBN). From the IBN, patient's physiological data are uploaded to the Health Server (HS). The prototype system was tested on people who volunteered for the project.

## 2. System Architecture

Fig 1 shows the overall WUHS three-tier layer system architecture. These layers are Biomedical Sensor Layer (BSL), Intelligent Base Node Layer (IBNL) and Health Server Layer (HSL), which consists of web application server and the database.

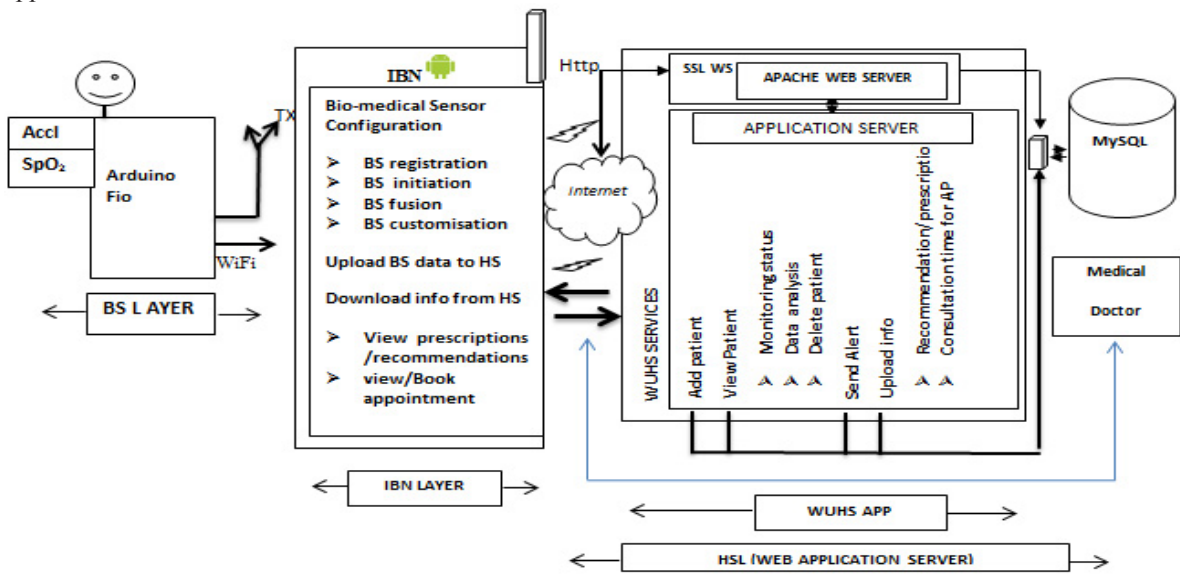

Fig. 1. WUHS System Architecture

The BSL consists of biomedical sensors implemented on arduino fio platforms. These sensors were developed to gather physiological data from patients and transmit the data through an attached WiFly radio over a Wi-Fi wireless network to the IBNL. The biomedical sensors are intelligent devices that can sense, sample, process and communicate physiological data. They also receive configuration instructions and responds to commands from the IBN, which is more superior in terms of intelligence [7].

The IBN in IBNL collects and processes physiological data received from biomedical sensors, it provides a Graphical User Interface (GUI) for patients to view their data in real time and uploads patient data to the Health Server (HS). In addition, it communicates with the HS via GPRS/Internet. The architecture uses an android smart phone with operating system (OS) 2.3v as the IBN. Android OS allows for real time data processing and higher processing power. Integrating an android smart phone with tri-axial accelerometer provides a way to determine mobility and Global Positioning System (GPS) for location determination. This makes it more suitable and a benefit to prototype WUHS. With services deployed on the IBNL, we can configure the biomedical sensors in terms of registration, initialization, fusion and customisation. The other services also deployed enable patients to upload physiological data, download recommendations or prescriptions information and view the medical doctor's work routine, if uploaded to the HS. This will enable patients' book medical appointments when needed.

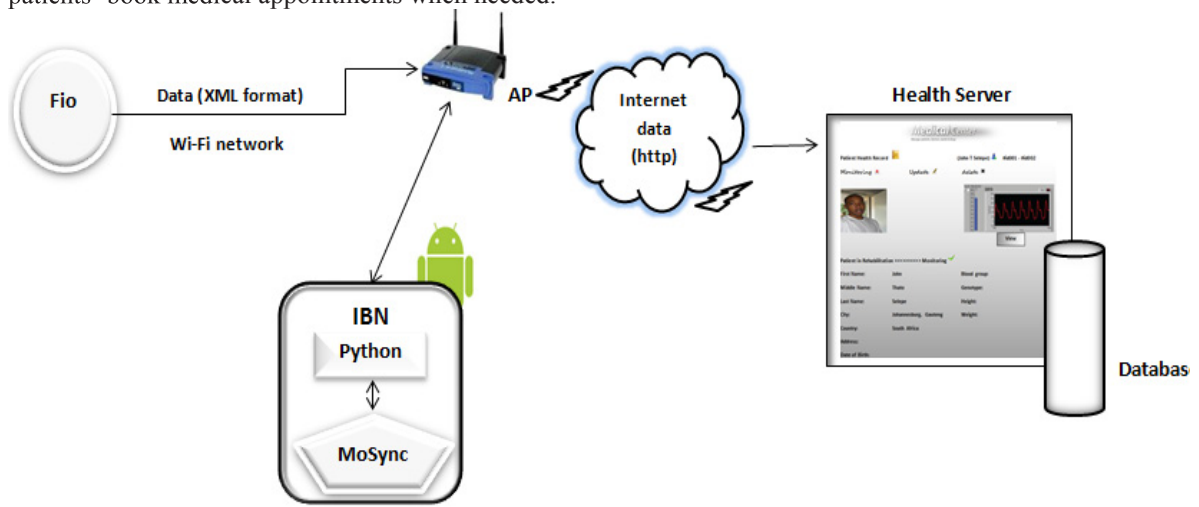

Fig. 2. Prototype WUHS data transmission process

The HS combines a web server and application server forming a Web Application Server (WAS), which can be accessed through a web browser. The developed WAS is deployed on Tomcat with Apache and MySQL database. On the application component, *Add patient* service registers and initializes a new patient record on the database. *View patient* service checks the existing patient medical record such as monitoring status or history and received data analysis. *Send alert* service creates an alert when data analysis results in a potential medical condition. With the *upload info* service, patient recommendations, prescriptions and medical doctor's consultation slots can be uploaded to the HS. Two main services offered at the web server component are *enterData* and *collectData*. The *enterData* service enables data to be entered into the database and *collectData* helps retrieve data. The medical doctor has to log in to access these services, which give the medical doctors access to the patient's record on the HS over any public network provided the correct authentication details are used.

## 3. System Implementation

The developed biomedical sensors have the capacity to collect large amount of physiological data and transmits these data continuously. Fig 2 shows the diagrammatical description of the WUHS data transmission process.

The accelerometer gives the continuous monitoring and transmission of the patient's angular force on the X, Y and Z axes. These movements are translated into motion states that correspond to the activities of patient in resting (ID - 1), walking (ID - 2), running (ID - 3) and falling (ID - 4) states. To carry out measurement test, the accelerometer is sampled at a frequency of 50Hz on all axes. Fig 3 shows the test measurements for the ID - 2 and ID - 4 states. We use AcclX, AcclY and AcclZ to respectively represent the front, side and vertical accelerations. Due to the earth's gravitational force, there is always an output of +g present if the patient is in a vertical position and vertical measurement AcclZ is taken. During test measurements, ID - 1 and ID - 2 showed acceleration movements patterns mostly on the X and Y axes while the Z axis showed less acceleration movement patterns. Measurement tests for ID - 3 and ID - 4 also showed acceleration movements patterns mostly on Z axis while X and Y axes showed less acceleration movement patterns.

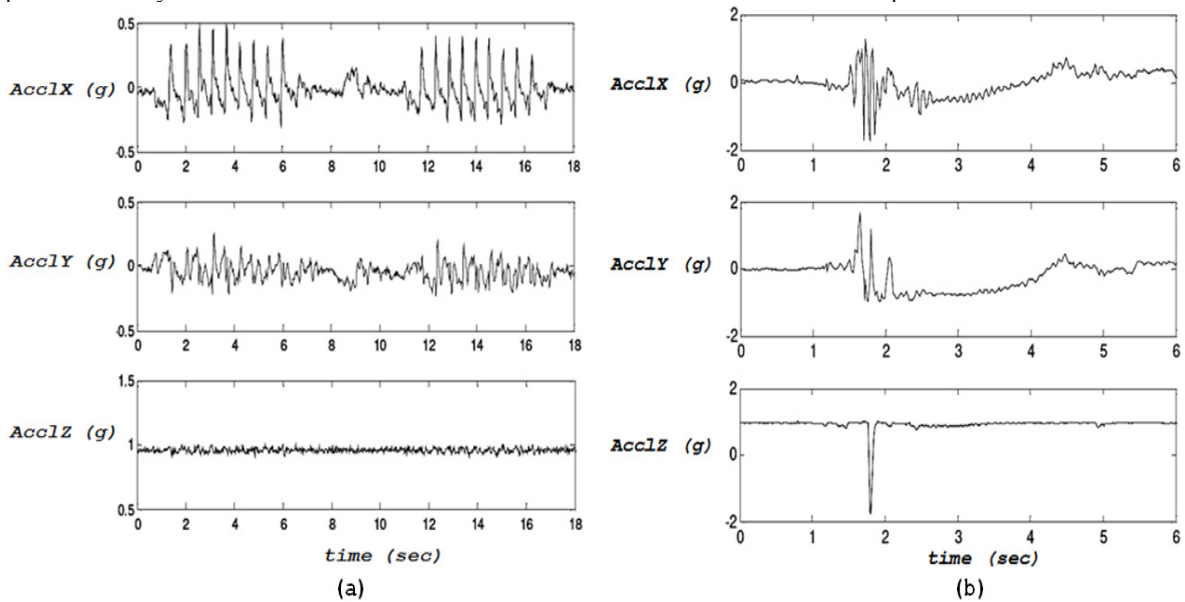

Fig 3. Accelerometer measurements showing (a) walking (ID – 2) and (b) falling (ID – 4)

The measurement results show different patient activities that generate different patterns in acceleration readings. With this, different human activities can be grouped through analysing the recorded acceleration data and warn against abnormal activities such as falling. Moreover, once an abnormal activity is detected through the acceleration reading, the health status can be verified with inputs from other sources, for example by checking the heart rate from a $SpO_2$ or heart activity from an ECG sensor.

The measurements of $SpO_2$ are based on Lambert Beer's law of spectral analysis that relates the concentration of absorbent in solution to amount of light transmitted through the solution [8]. Knowing the intensity, the path length and extinction co-efficient of a substance (here, oxyhemoglobin or reduced hemoglobin) at a particular wavelength, we determine oxygen saturation by measuring the light transmitted at two different wavelengths through the fingertip. Light spectrum at the red region around 660nm reduced hemoglobin (Hb) has higher extinction co-efficient compared to oxyhemoglobin (HbO$_2$). While at the near-infrared region of the light spectrum around 940nm, the extinction co-efficient of Hb is low compared to HbO$_2$. Using the differences in extinction co-efficient, we calculated for ratio R, which correlates to oxygen saturation [10].

The pulse oximeter test calculations report hearts rates in the range of 30 – 245bpm and SpO$_2$ values from 0-97%. The pulse oximeter performs all required calculations and transmits physiological data via a WiFly radio over the Wi-Fi network to the IBN. The Wi-Fi wireless network ensures efficient transmission of patient physiological data using deployed our TDMA schemes.

We have implemented a MoSync python application to run on the android smart phone. The application gathers data sent from the biomedical sensors. The IBN uses the python APIs (Application Program Interfaces) to manage the AP's processes such as setting up wireless connections to both biomedical sensors and HS. On receiving data from the biomedical sensors, the IBN uses comparison algorithm to compare data received with the last data sent. The IBN sends data to the HS only when there is a difference in data just received from the last data sent. This will help save cost for the patient by reducing the number of data transmissions made. The IBN also enables the patient view his/her medical status in real time on its GUI. The web application server mines data received and determines health risks using logistic regression. Functionality on the web application server is implemented in HTML/PHP. The medical doctors can view the patient's activities and SpO$_2$ health status in real time through a web application over a web browser. An abnormal condition triggers the alert signal, which is received by both patient and medical doctors and in high risk situations the GPRS on the android can be used to determine the patient's location.

The implementation of various off-the-shelf components and open source software (Python, PHP, MySQL and Apache Tomcat) helps to reduce the overall cost of WUHS. The power consumption and size for the major components in biomedical sensors and IBNL to give a cost evaluation as discussed [10].

## 4. Conclusion

The WUHS enables the monitoring of patients physiological data in real time, promising ubiquitous, yet an affordable and effective way to provide healthcare services. WUHS is one of the first that successfully integrates biomedical sensors with web services to monitor physiological data. The WUHS ensures interoperability between heterogeneous devices and technologies, data representations, scalability and reuse. The WUHS also ensures privacy, integrity and authentication protocols by using passwords and encryptions because privacy is seen as a fundamental right of humans and is a sensitive issue in healthcare applications. The important contributions of this paper are succinctly summarized as follows. The development of a system that achieves cost effective ubiquitous healthcare services by seamlessly integrating biomedical sensors. The developed system provides reliable and power efficient medical data transmission within the 802.11/Wi-Fi using TDMA schemes.